# Top-down fabrication of high-uniformity nanodiamonds by self-assembled block copolymer masks

Jiabao Zheng[1], Benjamin Lienhard[1], Gregory Doerk[2], Mircea Cotlet[2], Eric Bersin[1], Harrison Sejoon Kim[3], Young-Chul Byun[3], Chang-Yong Nam[2], Jiyoung Kim[3], Charles T. Black[2], Dirk Englund[1]

1 Department of Electrical Engineering and Computer Science, Massachusetts Institute of Technology, Cambridge, Massachusetts 02139, United States

2 Center for Functional Nanomaterials, Brookhaven National Laboratory, Upton, NY 11973, USA

3 Department of Materials Science and Engineering, The University of Texas at Dallas, 800 West Campbell Road, Richardson, Texas 75080, USA

## Abstract

Nanodiamonds hosting colour centres are a promising material platform for various quantum technologies. The fabrication of non-aggregated and uniformly-sized nanodiamonds with systematic integration of single quantum emitters has so far been lacking. Here, we present a top-down fabrication method to produce 30.0±5.4 nm uniformly-sized single-crystal nanodiamonds by block copolymer self-assembled nanomask patterning together with directional and isotropic reactive ion etching. We show detected emission from bright single nitrogen vacancy centres hosted in the fabricated nanodiamonds. The lithographically precise patterning of large areas of diamond by self-assembled masks and their release into uniformly sized nanodiamonds open up new possibilities for quantum information processing and sensing.

## 1 Introduction

Over the past two decades, solid-state defects have emerged as one of the leading systems for a wide variety of quantum technologies. Solid-state hosts such as diamond or silicon carbide are well studied, and a wide spectrum of fluorescing crystal defects have been identified and characterized[1–5]. In particular, the diamond nitrogen vacancy (NV) centre with its optically-addressable long-lived spin system is well suited for applications ranging from quantum networks[6,7] to quantum sensors[8]. Research interest has been growing considerably in developing other atomic emitters, such as the group-IV vacancy centres, possessing similar characteristics but with improved spectral stability[9–12].

Nanodiamonds (ND) hosting such colour centres are promising for various biological[13] and quantum[14–16] technologies, thanks in part to their compatibility with biologically active tissue and with common surface modification techniques[13,17]. Furthermore, they have been employed for laser trapping techniques[18–20] and scanning tip microscopy [21–23]. In comparison with bulk diamond, NDs offer nanometre-scale spatial positioning and compatibility with optical levitation. Typical commercialized techniques for ND fabrication include detonation, laser ablation, balling milling, high-pressure high-temperature (HPHT) growth, and chemical vapor deposition (CVD) growth[24]. However, these methods lack control over size or aggregation of the resulting NDs[14]. A reliable fabrication method of non-aggregated, uniform-sized, monocrystalline NDs with incorporated single colour centres is still challenging.

Control over the ND size uniformity is an important figure of merit for applications that involve building nano-hybrids[25], self-assembly processes[26,27], or integration with quantum emitters created via ion implantation[28]. Controlling the ND size requires precise tuning of the detonation, milling, or growth process and is usually followed by an extra ultracentrifugation step. Furthermore, the separation of individual NDs[29] is an important requirement to prepare ND samples for subsequent integration with photonic nanostructures[30] or biological tissue[31]. Traditional methods require stabilizing agents which can introduce contamination during the sample preparation[32]. Furthermore, control of doping concentration in NDs is necessary for applications such as nanodiamond electronics[33] and optical levitation of NDs[34]. Methods based on ion implantation[28] or detonation synthesis[35] are part of ongoing research.

Solutions to these fabrication challenges have been explored, and while progress has been made, many problems still remain. A fabrication approach for reactive ion etching of NDs using a sputtered gold mask with mean particle diameters of 50 nm has been demonstrated and resulted in NDs hosting implanted NVs with spin coherence times, $T_2$, exceeding 200 μs[36]. However, such sputtered gold masks suffer from significant non-uniformity, and the mechanical process utilized to release diamond nanopillars from the parent diamond causes loss of material. Another option to fabricate NDs is to use a CVD grown diamond membrane with a delta-doped NV layer[37]. Electron beam lithography (EBL) and plasma etching can then be used to control the ND size. This technique enabled the demonstration of NVs in NDs with 200 nm diameter on average and a $T_2$ exceeding 700 μs. However, the EBL fabrication of the mask is not scalable, and the resultant particles exceed the size requirements for many applications.

Here, we demonstrate large-scale parallel fabrication of non-aggregated, uniform-sized, monocrystalline NDs hosting single NV centres. The fabrication technique starts with CVD grown high-purity single-crystal diamond. It leverages the scalability of block copolymer (BCP) self-assembly combined with sequential infiltration synthesis to define nanometre-sized etching masks across an arbitrarily large diamond sample. A directional plasma etching step defines the dimensions of the NDs. An isotropic plasma etching step releases the NDs. We confirmed single-photon emission from single NVs hosted in the fabricated NDs. Statistics of the diameter of the released NDs indicate a mean diameter of about 30 nm with a variance of 5.4 nm.

# 2 Results

Our ND fabrication process starts with commercially available bulk monocrystalline diamond (Element Six) grown by a microwave-assisted CVD process. We use IIA optical grade monocrystalline diamond (size 3 x 3 x 0.5 $mm^3$) with a nitrogen concentration of less than 5 ppm (corresponding to ~ 12 nitrogen atoms for 30 nm diameter diamond spheres). We deposit a ~ 30 nm $SiO_2$ layer by plasma-enhanced chemical vapor deposition (PECVD) as an etch stop layer. As outlined in Figure 1, the fabrication process proceeds as follows: (a) The mask array in a hexagonal lattice configuration is produced by BCP self assembly (SA), followed by sequential infiltration synthesis (SIS)[38,39] to selectively load dots with $AlO_x$ to produce a hard etch mask that can withstand the subsequent reactive ion processing steps, see Supplemental Information (SI) for details. (b) Oxygen plasma reactive ion etching (RIE) removes the polymer, leaving only the $AlO_x$ dots array pattern. (c) Dry etching of $SiO_2$ using $SF_6$ and oxygen plasma transfers the hexagonal dot array pattern into the $SiO_2$ etch stop layer. (d) Directional oxygen plasma etching transfers this $SiO_2$ dot array pattern into the diamond, producing a hexagonal array of diamond pillars with a height tunable by the directional oxygen RIE duration, targeted here to 30 nm. (e) Atomic layer deposition (ALD) of $SiO_2$ protects the diamond pillar sidewalls. (f) Directional plasma etching of the $SiO_2$ using an $SF_6$ and $O_2$ gas mixture removes the $SiO_2$ layer at the non-sidewall surfaces. (g) Quasi-isotropic oxygen plasma etching[40] partially undercuts the bottom surface of the diamond pillars, see SI for details. At the end of this process, the ~ 30 nm sized diamond nanocrystals are nearly free-standing on ultrathin (~ 4 nm) pedestals, as shown in the close-up scanning electron micrograph (SEM) in Figure 2a. (h) Finally, hydrofluoric (HF) acid removes the $SiO_2$ masks and the passivation layers, leaving the diamond nanocrystal for characterization or subsequent releasing steps.

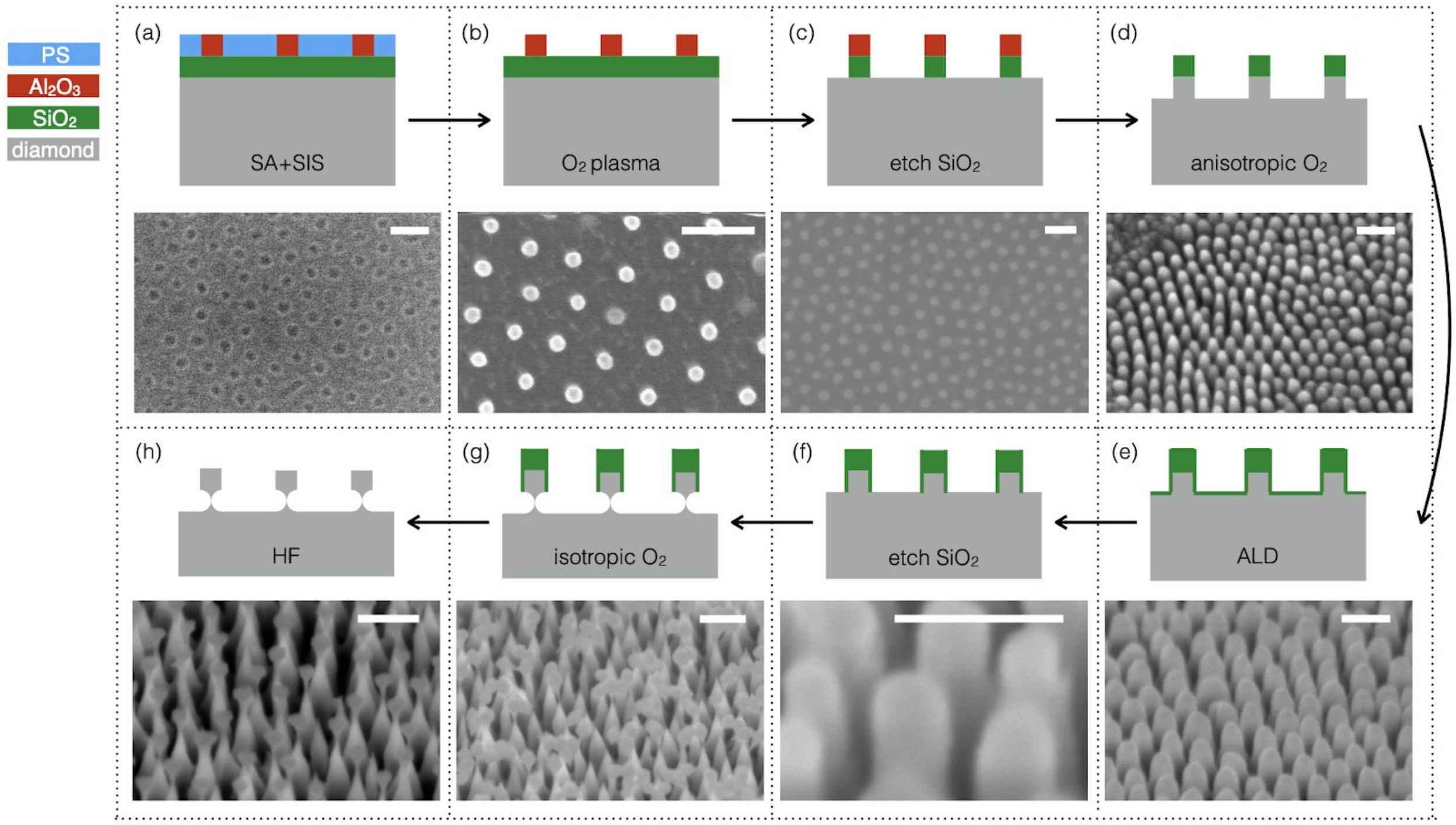

Figure 1: Fabrication process flow in 8 steps, with colour legend at upper left indicating the material of each layer. Each subfigure corresponds to one fabrication step, with schematic side-view diagram, and text showing the specific process carried out in the particular step, together with SEM images taken after that step, showing resultant device morphology. All scale bars indicate 100 nm. Note that images for steps (a), (b) and (c) are top view, and images

for other steps were obtained with samples tilted at 45 degrees. (a) Polished bulk single crystal diamond, coated with ~ 30 nm $SiO_2$ layer using PECVD, with $AlO_x$ hexagonal dot array patterned by BCP film using spin coating, SA and SIS. (b) Gentle $O_2$ plasma treatment removes the polymer content in the BCP film, leaving $AlO_x$ hexagonal dot array pattern. (c) Directional plasma etching of $SiO_2$, transferring the dot array pattern to $SiO_2$ layer. (d) directional $O_2$ plasma etching of diamond, transferring the dot array pattern to bulk diamond, forming diamond pillars arranged in hexagonal array. (e) ALD of ~ 2 nm $SiO_2$ coating to protect the sidewall of etched diamond pillars. (f) A short directional plasma etch removes $SiO_2$ from bottom surfaces of the open gaps between diamond pillars. (g) The sample is exposed to a quasi-isotropic $O_2$ plasma to undercut the diamond pillars. (h) The bulk diamond is immersed in HF acid to remove the residue $SiO_2$ layer.

**ND Size Distribution:** To characterize the fabricated NDs, we released the NDs on commercially available glass coverslips. We then performed large area SEM imaging of the NDs, as exemplified in Figure 2b. Customized image processing detects single NDs on the SEM images and estimates the dimensions based on the image contrast. We evaluated over 100 NDs and visually confirmed the boundary detection results. The resultant ND diameter statistics are fitted to a Gaussian distribution curve, as shown in Figure 2c in red. The Gaussian fit yields a mean diameter of 30.0 nm with the 95% confidence interval from 28.8 nm to 31.2 nm, and a variance of 5.4 nm with the 95% confidence interval from 4.7 nm to 6.4 nm.

**Optical Characterization:** The optical measurements were performed in a home-built confocal setup with 532 nm laser excitation and an excitation power of ~ 3 mW after the objective. The collection path can either be directed through a 690/40 nm notch filter to a single-photon avalanche photodiode (SPAD), or through a 561 nm long pass filter to an optical spectrometer.

Figure 2d shows a confocal scanning map of NDs released on a coverslip. The scan reveals a single bright photoluminescent spot; the corresponding spectrum collected from the bright spot, as shown in Figure 2e, indicates a room temperature NV emission spectrum in its negative charge state with the characteristic zero phonon line (ZPL) at 636.5 nm.

Second-order autocorrelation measurements were conducted with a home built confocal setup with a 594 nm laser at ~ 200 μW excitation power on the sample to excite the $NV^-$ charge state, and a 635 nm long pass filter followed by a Nikon oil immersion objective with numerical aperture (NA) of 1.3. The second-order autocorrelation measurements ($g^{(2)}(0)$ ~0.08) in Figure 2f indicate emission from a single NV centre with an emission of up to 80,000 single photons per second, recorded with SPADs.

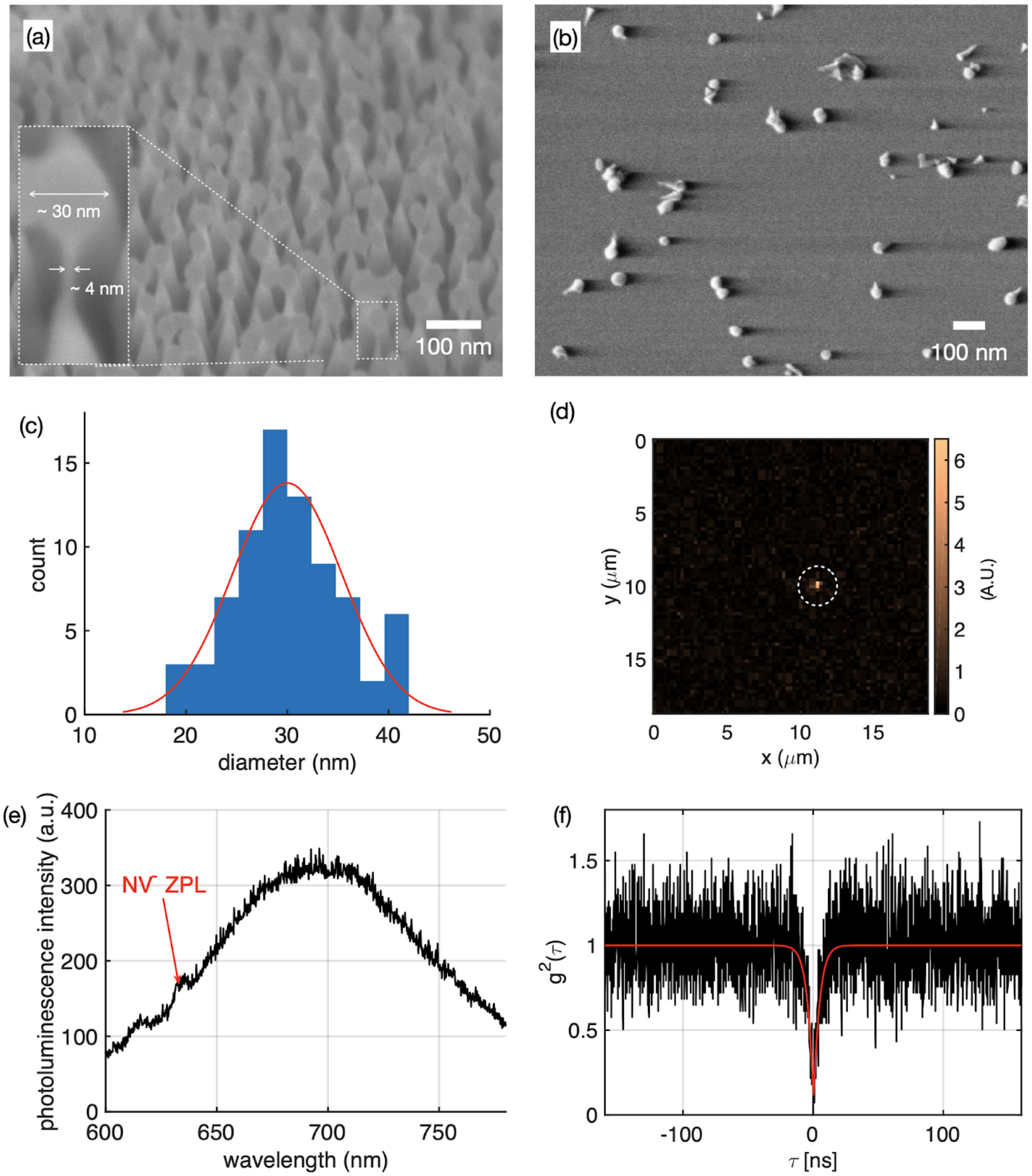

Figure 2: (a) SEM of fabricated NDs before release, featuring ~ 30 nm sized NDs sitting on ~ 4 nm diamond pedestals. (b) SEM of NDs released on glass substrate. (c) Particle size distribution histogram of NDs, determined from large scale SEM imaging and subsequent image processing to detect the size of the NDs in the SEM image. The red curve shows a Gaussian fit to the histogram with a mean diameter of 30 nm and a variance of 5.4 nm. (d) Optical confocal scanning image of the NDs released on glass substrate. The bright spot (white dashed circle) indicates an ND hosting a single NV colour centre. (e) Photoluminescence

spectrum collected from the bright spot shown in (d), with the characteristic NV zero phonon line (ZPL) at 636.5 nm and a broad phonon sideband. (f) Second-order autocorrelation histogram of the collected emission from the bright spot in (d), showing $g^2(0)$ of 0.08 indicating single photon emission ($g^2(0)<0.5$).

# 3 Discussion

We presented a top-down fabrication technique to produce non-aggregated monocrystalline NDs with a uniform size. The presented method enables the fabrication of NDs with a mean diameter of 30 nm and a variance of 5.4 nm. The process is performed on a IIA grade bulk diamond with a known nitrogen dopant concentration of 5 ppm. Second-order autocorrelation measurements confirm the presence of single quantum emitters in the ND.

This method can be applied to bulk diamond, diamond with tailored dopant properties, or isotopically purified diamond to produce NDs with the same dimensional properties[41,42]. Doped NDs are attractive materials for the study of medical oncology[43], electrochemistry[44,45], diamond electronics[46], and superconductivity[47,48]. Diamond enriched with $^{12}C$ can enable increased coherence times of hosted spin systems due to the limited number of interfering spin impurities in the host material[42].

The produced ND dimensions are determined by the mask patterning technique and the vertical etching step, shown in Figure 1(d). This scalable mask patterning and etching technique is compatible for producing NDs with a wide range of dimensions and aspect ratios. The size of the dot masks obtained by BCP SA followed by SIS can be tuned from ~ 10 to ~ 30 nm in diameter by changing the SIS cycle number[39,49], nanoimprint lithography can be used[50] for ~ 30 - 1000 nm in size, and traditional optical lithography is applicable for mask sizes beyond 1 μm. The vertical height of the NDs can be independently controlled by the dwell time of the vertical etching in the processing step shown in Figure 1(d).

Upon completion of the fabrication process, the NDs are non-aggregated, since the nanomask defined in the BCP SA process naturally separates one domain from another. The quasi-isotropic plasma undercut process exposes a large surface area and allows further surface treatments of the NDs. Chemical treatments such as surface modifications to minimize aggregation[27,32] or surface functionalization for bio-binding[51] are important prerequisites for a wide range of research processes and topics[52–54]. Such surface treatments are compatible with the underlying ND fabrication method. Furthermore, different surface terminations can change the quantum properties of NV centres[55], such as a nitrogen plasma treatment[56], or can stabilize the charge state of NV centres[57], such as a fluorine-based or oxygen-based plasma treatment.

The exposed surfaces of the resulting NDs provide an opportunity to grow conformal coatings for protection or functionalization before harvesting. For example, ALD of isotopically purified $SiO_2$ can be used to “package” the NDs with fewer spin impurities to improve the spin coherence of the hosted NVs. Such an ALD process can be achieved with precursors such as silane[58], for which isotropically enriched sources are commercially available thanks to the well-developed silicon growth process[59]. The oxidant for the $SiO_2$ ALD process can originate from naturally occurring oxygen. Naturally occurring oxygen has an abundance of 99.76% nuclear-spin-free $^{16}O$, higher than the naturally occurring 98.93% abundance of nuclear-spin-free $^{12}C$ in diamond.

The estimated production yield can be scaled up using a larger area and iterating in the next depth layer of the parent diamond. This process promises large-area fabrication of NDs with size and shape uniformity, thanks to the scalability of the BCP SA process. As demonstrated in this work, the ND production yield scales with the parent diamond area size. The 70 nm pitch used in the BCP SA nanomask template translates to ~ $2 \times 10^8$ NDs with ~ 30 nm in diameter for every 1x1 $mm^2$ diamond area. This corresponds to a production yield of ~ 9.89 $ng/mm^2$ measured in weight per unit area of the parent bulk diamond. However, the overall mass of nanodiamonds produced in this planar patterning method is of course much lower than it would be for 3D fabrication methods, such as ball-milling, though these 3D fabrication methods are unable to achieve the same size uniformity.[60]

The potential impact of the presented ND fabrication technique goes beyond size uniformity and non-aggregation. The quasi-isotropic plasma undercut may induce less strain on the NDs compared to other methods for top-down fabrication of nanodiamonds[36,37], though further studies are required. It would also be relevant to study the spin coherence times of group IV vacancy centres hosted in NDs produced by this process[11]. The spin coherence time $T_2$ of the silicon vacancy (SiV) centre is limited by the phonon-induced relaxation between two orbital ground states that are split by ~ 47 GHz [61]. NDs with diameters below half the corresponding phononic wavelength (~ 127 nm) may suppress the coupling of these phonon modes to prolong the SiV spin coherence times.[62] At the same time, it is important to minimize stress / strain in the NDs to maintain the defect centre's inversion symmetry such that their optical properties are preserved[63].

# 4 Conclusions

We realized a wide-area fabrication process in a top-down approach to produce non-aggregate monocrystalline NDs with high size uniformity. Additional optical characterization confirmed the existence of single NVs hosted in the NDs by probing their photoluminescence spectra and photon statistics. The fabrication technique applies to a broad range of engineered bulk diamond, including for example isotopically purified substrates. $^{12}C$ enriched diamond hosted spin systems have been shown to greatly extend the NV electron spin coherence time[36,37].

# 5 Supplementary Information

**RIE Mask Creation by BCP Self-Assembly:** To prepare the sample surface for BCP self assembly, we first clean the $SiO_2$ layer by oxygen plasma etching (March Plasma CS1701) for 1 minute at 20 W power with 100 mTorr pressure. To promote vertical orientation of self-assembled BCP domains, a "neutral" brush consisting of a hydroxyl-terminated polystyrene (PS) and poly(methyl methacrylate) (PMMA) random copolymer (PS-r-PMMA-OH, 61 mol. % styrene, determined by $^{13}C$ NMR, Mn = 9.2 kg/mol and Mw/Mn = 1.35, determined by gel permeation chromatography relative to PS standards), provided by The Dow Chemical Company (10.1063/1.5000965), is grafted to the $SiO_2$ surface by a dehydration reaction facilitated by baking the sample on a hot plate for 5 minutes at 250 °C in a nitrogen enriched environment. Ungrafted brush polymer is removed by subsequent spin-rinsing in propylene glycol monomethyl ether acetate (PGMEA). We then spin coat (3000 rpm) the BCP solution, 1% (w/w) in toluene, which is based on a cylindrical-phase polystyrene-b-poly(methyl methacrylate) (PS-b-PMMA) BCP (177 kg/mol, PS:PMMA = 131:46, Mw/Mn = 1.10, purchased from Polymer Source). The sample is thermally annealed on a hot plate for 20 minutes at 250°C in a $N_2$ enriched environment[64,65] to facilitate self-assembly. This self assembly process produces a hexagonal array of PMMA dots in a PS matrix as the thermodynamically favorable arrangement of the system, where the mean dot diameter is ~ 30 nm and the lattice period of the hexagonal array is ~ 70 nm.

Following self-assembly, SIS is used to selectively load the PMMA polymer domain with $AlO_x$ via adsorption of trimethylaluminum (TMA) and $H_2O$ sequentially for 4 cycles (100s dwell/purge times) in a commercial atomic layer deposition reactor (Cambridge Nanotech Savannah S100). After oxygen plasma ashing (March Plasma CS1701, 20 W power, 100 mTorr, 5 minutes) to remove both the PS and PMMA polymers, a hexagonal array pattern of $AlO_x$ dots remains and used as a hard masks for etching into $SiO_2$ layer and diamond. The form of the hexagonal array of $AlO_x$ dots with ~ 30 nm in size and ~ 70 nm in pitch is shown in the SEM image in Figure 1(b). We then use $SF_6$ and oxygen plasma dry etching to transfer the hexagonal array dots pattern to the $SiO_2$ layer, as indicated in Figure 1(c). A directional oxygen plasma etching is performed to etch the diamond down to ~ 30 nm, which defines the height of the diamond nanocrystals. The resultant structures are diamond pillars with ~ 30 nm in height, which are shown in Figure 1(d) with side-view schematic and SEM images taken with the sample tilted by 45 degrees.

**Isotropic Dry Etching for Undercutting Nanodiamond Pillars:** With the size of the diamond nanocrystals defined by the size of the $AlO_x$ dots and dwell time of the directional oxygen etching, it is necessary to undercut and release the diamond nanocrystals to complete the process. Undercutting nanostructures in fabrication has been realized by either homoepitaxial structure with sacrificial layers beneath the device layer, or engineered plasma etching with a Faraday cage[66,67] or zero RF driving power[40]. Here, we pursue the idea of quasi-isotropic etching for the release of diamond nanocrystals. Sidewall passivation with ~ 2 nm of conformal coating of ALD $SiO_2$ is performed to protect the diamond from the subsequent quasi-isotropic oxygen plasma etching, as shown in Figure 1(e). The ALD was performed in a viscous flow reactor using tris[dimethylamino]silane (3DMAS, 99.999%, Sigma-Aldrich) and ozone ($O_3$) as Si source and oxidant, respectively, at the process temperature of 250 °C. Growth thickness per cycle for the process was confirmed to be 1.3 Å/cycle. We used an exposure-mode ALD in which 1 ALD cycle consists of: 0.05 s dosing of 3DMAS, 28 s waiting under static vacuum, 13 s chamber purging (for removing excess precursors and reaction byproducts) under 100 sccm nitrogen flow, 0.2 s dosing of $O_3$, 7 s waiting under static vacuum, and, finally, 13 s chamber purging under 100 sccm nitrogen flow. Total 15 ALD cycles were applied to yield approximately 2 nm of $SiO_2$ conformal coating on etched diamond pillars. We then use a directional etching with a mixture of $SF_6$ and oxygen to selectively remove the $SiO_2$ on the non-sidewall surfaces, which opens up the bottom surfaces for the subsequent quasi-isotropic oxygen plasma etching of the diamond. We heat our samples at 90°C and slightly drive the plasma with ~ 5 W of biasing power, which gives a horizontal etch rate of ~ 0.5 nm/min at the sidewall surfaces with a vertical etch rate of ~ 2.5 nm/min. This undercut step enables the control over the size of the pedestal that the diamond nanocrystals sit on by timing the duration of the quasi-isotropic etching step. The diamond nanoparticles are undercut to nearly free-standing, so that they can be easily harvested. The final step is a HF acid immersion to remove the residue $SiO_2$ and $AlO_x$, after which we release the NDs on a ~130 μm thick glass coverslip for optical characterization. The release step is done by bringing the surface of the bulk IIA diamond in direct contact with the glass coverslip, and some NDs are attached to the surface of the the glass coverslip, possibly due to electrostatic binding.

# Acknowledgements

The conducted research was supported in part by the Army Research office MURI biological transduction program and NSF Center for Integrated Quantum Materials. Research carried out in part at the Center for Functional Nanomaterials, Brookhaven National Laboratory, which is supported by the U.S. Department of Energy, Office of Basic Energy Sciences, under Contract No. DE-SC0012704. J.Z. was supported in part by Master Dynamic, Inc.

# Author contribution statement

J.Z. and G.D. developed the fabrication technique. H.S.K., Y-C.B. and J.K. conducted the atomic layer deposition of silicon dioxide. B.L., J.Z. and M.C. characterized the fabricated sample. C-Y.N. and G.D. trained J.Z. on the cleanroom and nanofabrication facilities. J.Z. and B.L. wrote the manuscript in consultation with E.B. and D.E.. C.T.B. and D.E. supervised the project. All the authors commented and revised the manuscript.

**Competing financial interests:** The authors declare no competing financial interests.

**Data availability:** The data that support the findings of this study are available upon reasonable request.